# Ion Transport Through Cell Membrane Channels

Jan Gomułkiewicz[1], Jacek Miękisz[2] , and Stanisław Miękisz[3]

[1] Institute of Physics,  Wrocław Technical University;

[2] Institute of Applied Mathematics and Mechanics, Warsaw University;

[3] Department of Biophysics, Medical University of  Wrocław

**Abstract.** We discuss various models of ion transport through cell membrane channels. Recent experimental data shows that sizes of ion channels are compared to those of ions and that only few ions may be simultaneously in any single channel. Theoretical description of ion transport in such channels should therefore take into account interactions between ions and between ions and channel proteins. This is not satisfied by macroscopic continuum models based on Poisson-Nernst-Planck equations. More realistic descriptions of ion transport are offered by microscopic Brownian and molecular dynamics. One should also take into account a dynamical character of the channel structure. This is not yet addressed in the literature.

## Introduction

Every living cell is an open system. Continuous transfer of energy and mass between cells and their surroundings constitutes a necessary condition for a sustained life. Cell membranes, which ensure the autonomy of separated compartments, should be endowed by mechanisms of a selective transport of substances indispensable for the life of a cell. In particular, a fundamental phenomenon is a transport of ions through cell membranes which ensures that the ion content of a cell is different from the one outside the cell.

Cell membranes, due to their structure and a chemical composition (a two-lipid layer with immersed molecules of integral proteins) are characterized by a very low average relative electrical permittivity $\varepsilon_r$ (also called dielectric constant) about 2,as compared to a relative

---

permittivity of electrolyte water solutions in their surrounding – about 80. The Born energy (the energy required to move a ion from the outside solution to the hydrophobic interior of a membrane) corresponding to such values of permittivity is about 300kJ/mol. This rather high Born energy should imply a very low ion permeability of a membrane (in particular for such important ions as Na, K, Ca, and Cl) – the value lower by several orders of magnitude than observed ones. In evolutionary processes, structures and mechanisms have been formed in membranes, which lower locally an energy barrier for  penetrating ions. Such structures consist of molecules of integral proteins whose hydrophobic parts interact with two-lipid layers and polar hydrophilic parts form: a) relatively wide non-selective hydrated pores which penetrate membranes, b) specific ion channels, often endowed with special structural elements which form gates sensitive to an electric field, chemical ligands, or the mechanical stress, c) ion-binding centers (on one or both sides of a membrane) called carriers or transporters which interact with ions and transport them to the other side of the membrane where ion-carrier complexes dissociate. Such transport can use the energy obtained by metabolic reactions (mainly ATP hydrolysis) – then it is called the active transport, or an interior energy – in this case it is called the facilitated diffusion.

All above mentioned transport mechanisms are essential for cell homeostasis, that is for securing the content of the interior of a cell, its volume and an electric voltage of the membrane (in electrophysiology and biophysics it is called the membrane potential). They are also essential (to a certain degree) for the phenomenon of excitability of cells, for which an important role is played by ion channels. Investigations in this area are carried in many scientific centers. Huge experimental data has been collected and  various  theories proposed which describe ion transport in cell membranes. Despite this, our knowledge is still not sufficient to explain transport mechanisms and to provide its full description.

**Ion channels**

In the late forties and early fifties of the last century, Hodgkin and Huxley in the collaboration with Katz (Hodgkin & Katz 1949; Hodgkin & Huxley 1952; Hodgkin et al. 1952; Huxley 2002; Moves 1984) worked out their phenomenological theory of nerve impulses und put up a hypothesis that transport of potassium and sodium ions in excitable biological membranes takes places in selective ionic paths, different for different ions. These paths, besides a high selectivity, displayed the dependence of the conductivity on the membrane voltage. Although, such paths have not been called ion channels in these papers, yet a widely accepted hypothesis was formulated that ions penetrate membranes through specific ion channels made of proteins. Channel properties as well as transport mechanisms were deducted from macroscopic measurements. Particularly useful was the *voltage-clamp* method which consists of registering electric currents through a certain macroscopic surface of a membrane for fixed values of the membrane potential. A schematic set up for such measurements is presented in Fig.1 .

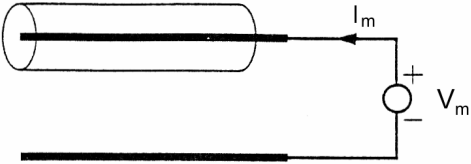

Fig. 1. Schematic set up for measuring ion currents in axons by the *voltage-clamp* method.

The registered current through the membrane is the sum of ion currents and the capacity current (a current of charging the membrane's electrical capacity). The second current vanishes with a time constant specified by the membrane's capacitance and the electrical resistance of the measuring circuit. The total current $I_m$ is written as

$$I_m = I_c + \sum_i I_i \, , \qquad\qquad (1)$$

where $I_c$ is the capacity current and $I_i$ are currents carried by respective types of ions (mainly sodium and potassium currents but also a chloride current one which is a main ingredient of the leakage current for which the electrical conductance is independent of the membrane potential). Using toxins blocking specifically particular type of ions, one can decompose experimentally the total current into different ion currents. This is illustrated in Fig.2.

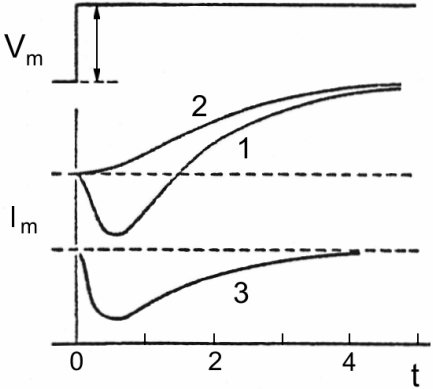

Fig. 2. Time dependence of ion currents for a fixed membrane potential.

After applying to a membrane a fixed voltage $V_m$, the total current is given by the curve 1. (the capacity current is not seen in Fig.2 due to its very low time constant). Then using a specific blocker, for example for sodium ions, one registers the total current diminished by its sodium component - line 2. Line 3 corresponds to the difference between values of currents given by lines 1 and 2 (assuming that the leakage current is negligibly small) and therefore describes the current crossing through all sodium channels on the membrane surface between measuring electrodes. There are many channels on this surface and therefore their individual properties can be only deducted from current-voltage characteristics obtained in concrete experimental conditions. One can read about the *voltage-clump* method for example in (Weiss 1996). The *patch-clump* worked out by Neher and Sakmann (1976) make possible measuring currents through individual channels. In this method, the ending of a glass pipette (of the diameter of 1μm and the resistance of the contact of $10^9$ ohms) is attached to the membrane. One can register electric currents through the surface adhered to the pipette for fixed voltage between measuring electrodes. There should be only one ion channel located on such a small surface and therefore one can obtain transport characteristics of individual channels. A schematic set-up of this method is presented in Fig. 3.

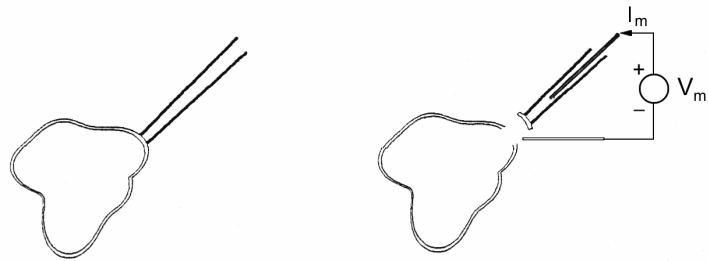

Fig. 3. Schematic set-up for measuring ion currents by the *patch- clamp* method .

We will not discuss details of this method which can be found in the very good monograph of Sakmann and Neher (1995).

Investigations using the *patch-clump* method confirmed a high selectivity of ion channels. Transport velocity of ions, obtained from these measurements, equals to about $10^7$ ions per second (a number of ions crossing a channel during one second) appeared to be close to values observed in the diffusion in water solutions of electrolytes with the thickness compared to that of cell membranes. Such high velocity of ions in membranes indicates that the transport mechanism cannot be of a carrier type which is the case in the active transport realized by ion pumps or in the passive exchange transport (for example realized in the case of anions by band 3 protein in the erythrocyte membrane). Channels in excitable membranes are highly selective for univalent kations ($Na^+$ and $K^+$). Such selectivity cannot be therefore the effect of electrostatic interactions between ions and charges of the inner surface of the channel. It results from interactions with chemical residues of channel proteins directed towards the interior of the channel (Beckstein & Sansom 2004; Gouaux & MacKinnon 2005; MacKinnon 2003; Miller 2000; Noskov et al. 2004a; Noskov & Roux 2006; Zhou & MacKinnon 2003). Biochemical studies tell us which proteins form particular channels. We know their aminoacid sequences and ternary and quaternary structures. It is known which parts of channel proteins are responsible for hydrated pores, which parts form a filter responsible for the channel selectivity and which ones play the role of a voltage sensor which can change the state of a voltage-dependent channel from the conductive to the non-conductive one (and vice versa). Bibliography discussing these issues is immense and we will not cite it here but rather refer readers to the new edition of an excellent monograph (Hille 2001) and review papers (Sansom et al. 2002; Tombola et al. 2006; Yesylevskyy & Kharkyanen 2004). A new idea on this matter is presented in MacKinnon's papers ( Lee et al. 2005; Schmidt et al. 2006).

Scientists who worked out theoretical description of ion transport in open channels had to base their models on biochemical data and experimentally obtained channel transport characteristics. Until recently we had lacked directly obtained channel images. First reports of such images appeared in the end of nineties of the last century. There have been obtained X-rays images of potassium and chlorine channels, an acetylocholin receptor, and water channels (aquaporins). Particularly important is a paper (Doyle at al. 1998) whose authors were able to crystallize the protein of a potassium channel, KcsA, from the membrane of the bacteria *Streptomyces lividans,* and obtained its three-dimensional X-ray image with 0,32 nm resolution. Results obtained in this paper were confirmed in (Morais-Cabral et al. 2001, Zhou et al. 2001) with images with 0,2 nm resolution. Despite the fact that the KcsA channel is not voltage-dependent and its image corresponds to the non-conductive state, it has become the base for constructing realistic models of ion channels (mainly potassium ones) (Sansom et al. 2002; Tieleman et al. 2001, and the literature cited therein).

The general scheme of the potassium channel following from the above papers is presented in Fig. 4.

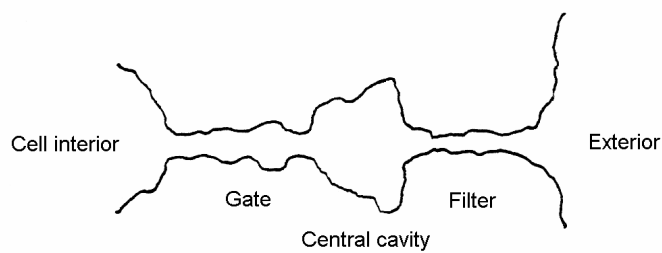

Fig.4. Schematic cross-section of a ion channel.

One can see that the channel cross-section changes along its axis. In the channel part directed toward the outside of the cell there is located a filter responsible for channel selectivity (of the length of 1,2 nm length and the diameter of 0,3 nm). In the middle part there is a relatively wide cavity of the length of about 1,0 nm which is capable of containing tens of water molecules. In the part of the channel directed toward the interior of the cell (of the length of about 2,0 nm) there are subunits of the protein channels responsible for the opening and closing the channel (a channel gate). In the closed channel, the smallest diameter of this part of the channel is 0,24 nm, whereas the diameter of $K^+$ ion is equal to 0,26 nm. The surface of this part of the channel is hydrophobic. The above data come from (Chung & Kuyucak 2002). It is seen from the X-ray image of the channel that there can be at most two potassium ions in the filter (separated by a water molecule). In the middle cavity of the channel there can be a third potassium ion. Such distribution of ions in a channel is confirmed by Brownian dynamics (Chung et al. 2002), and molecular dynamics (Allen et al. 1999; Burykin et al. 2002; Sansom et al. 2002, and the literature cited therein).

Detail studies of channel proteins indicate that a filter part is the same in all potassium channels (Hille 2001; LeMasurier et al. 2001; Morais-Cabral et al. 2001). It is formed from segments of polypeptide chains (two or four subunits of a channel protein) with the amino acid sequence TVGYG. One can distinguish four centers (S1, S2, S3, and S4) in which oxygen atoms of the carbonyl residues exactly correspond to the coordination bond of potassium ions and can substitute oxygen atoms of water molecules around the hydrated $K^+$ ion (Miller 2000). X-ray studies (Zhou et al. 2001) and molecular dynamics (Berneche & Roux 2001) indicate that there exists another center (S0) in the exterior of the channel entrance domain. This is schematically illustrated in Fig.5.

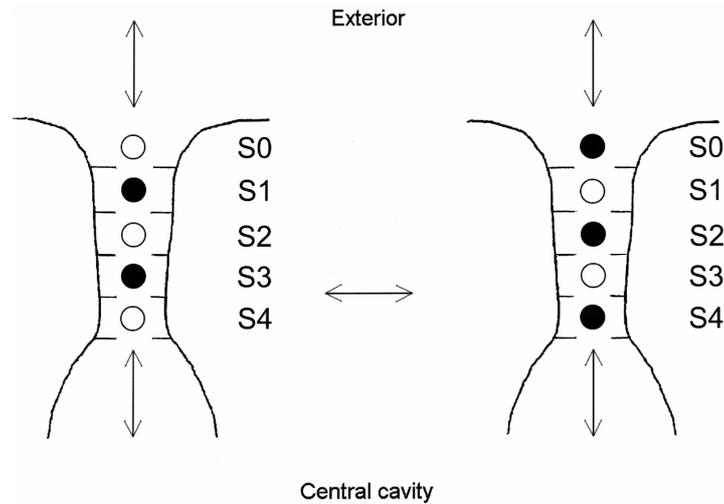

Fig.5. Schematic distribution of K$^+$ ions ( ● ) and water molecules ( ○ ) in the channel filter.

     The presence of such filter in all potassium channels allows the construction of a uniform theoretical description of the ion transport in all open potassium channels. It was shown in (Corry et al. 2001) that such description can also be used in calcium channels.

**Ion transport in channels**

In describing ion transport in membrane channels, one has to take into account both geometrical sizes of a channel and types of chemical residues of proteins on the channel surface. One cannot forget that an image of a ion channel obtained from structural studies is a static one (averaged over a certain time), whereas the real structure varies in time. Uptill now, all theories of ion transport (with the exception of the molecular dynamics) (Allen et al. 2004; Allen et al. 2006; Bastug et al. 2006; Beckstein & Sansom 2004; Chung & Tieleman 2006; Sands et al. 2006) were not taking this important fact into consideration.

     A continual electro-diffusion theory, proposed by Goldman (1943) and then developed in many papers has been and still is often used to describe the ion transport in open channels. We present foundations of such description in Appendix A. A comprehensive bibliography can be found in monographs (Hille 2001; Sten-Knudsen 2002). It is a mean-field theory. It can only be used to describe ion transport in channels of a sufficiently large diameter and for solutions of dilution ensuring that ions do not interact with themselves. In such channels, concentration of ions can be described as a statistical quantity. Electrical permeability and the diffusion coefficient for ions inside the channel are equal to the corresponding values in the solution adjacent to the membrane. However, this is not true for channels presented above. Eisenberg and his coworkers (Chen & Eisenberg 1993a; Chen & Eisenberg 1993b; Eisenberg

1998; Eisenberg 2000a; Eisenberg 2000b; Eisenberg 2003; Gillespie et al. 2002; Nonner et al. 1999; Schutz et al. 2001) and Kurnikova with her coworkers (Cardenas et al. 2000; Graf et al. 2004; Kurnikova et al. 1999; Mamonov et al. 2003) developed a three-dimensional electro-diffusion theory based on the Nernst-Planck equation and the Poisson equation for the potential of electrostatic interactions. This theory, called PNP (Poisson-Nernst-Planck) theory, is still a mean-field theory and as it was shown in (Corry et al. 2000; Moy et al. 2000), it cannot be used in channels of dimensions described above. It can be applied to channels of dimensions exceeding twice the value of the Debye radius. The PNP theory cannot be used to describe the transport in channels discussed here because the average number of ions observed in such channels is compared to the size of fluctuations and therefore the concept of concentration ceases to have a sense. This problem was pointed out in (Miller 1999; Schutz et al. 2001). However, this was not taken into account in proposed theories and in the case of the second paper, this problem was bypassed by averaging over a long time in the manner which is not completely understood. In a recent paper, the Eisenberg group generalizes the PNP equations by adding to the chemical potential of ions, an "excess" responsible for interactions between ions and non-electrostatic interactions between ions and the surface of a channel (Gillespie & Eisenberg 2002; Gillespie et al. 2002).

The PNP theory fails to explain an experimentally observed saturation of the ion flux as a function of the concentration of ions in the solution adjacent to the membrane for the fixed membrane potential. According to the PNP theory, this dependence should be linear.

It is worth to point out once again that using the continuous description of the ion transport in channels with atomic dimensions is inherently inappropriate. Macroscopic parameters of channels such as the diffusion coefficient (equivalently ions mobility), concentration and electric permeability, taken from continual theories cannot be rationally justified. This is confirmed by Monte-Carlo simulations of a double electric layer in 2 nm nano-pores (Yang & Yiacoumil 2002; Yang et al. 2002).

The above remarks about continual description of ion transport in open channels, based on the PNP theory suggest that it should not be used for channels with dimensions observed experimentally. Accidental agreement of this theory with an experimental data can follow from cancellation effects of assuming wrong channel parameters and wrong assumptions about the physics of channels (Corry et al. 2001).

A fundamental model aspiring to provide a realistic description of ion transport in channels of excitable membranes is the molecular dynamics (MD) (Allen et al. 1999; Allen et

al. 2006; Berneche & Roux 2001; Chung et al. 2002; Levitt 1999; Nadler 2002; Roux 2005; Roux et al. 2004; Chung & Tieleman 2006). Ions, molecules of water and of channel proteins are treated as individual objects. Newton equations of motion of ions interacting with other ions, water molecules and the surface of a channel are solved. Unfortunately, the computer time needed to solve these equations and to calculate properties of channels (like their conductivity) is so far prohibitively large (years for the fastest computers). In such situation, even if one could have a realistic model of channels and a proper physical theory describing interactions between molecules in a channel, then limitations in computer power prevent us from using molecular dynamics to describe the ion transport.

A model which is less fundamental but still describing motion of individual ions is the Brownian dynamics (BD). In order to reduce the number of equations, the force acting on a given ion (originated from water molecules and the surface of a channel) is decomposed into the sum of a deterministic friction force and a random force (a white noise) with the zero average. In this way we pass from deterministic Newton equations to stochastic Langevin equations,

$$m_j \frac{dv_j}{dt} = -m_j \gamma_j v_j + F_r(t) + q_j E + F_s , \qquad (2)$$

where $m_j$, $q_j$, $v_j$, are respectively a mass, a charge, and a velocity of the j-th ion, E is the electric field. The friction, $m_j\gamma_j v_j$, (where $\gamma_j$ is the friction coefficient per unit mass) and a stochastic force, $F_r$, are results of random collisions of ions with water molecules and the channel surface. $F_s$ is the force of a short-range non-electrostatic interaction between the ion and the channel. The electric field responsible for the force exerted on ions, is computed numerically from the Poisson equation in the form

$$\nabla\big[\varepsilon(\bar{r})\nabla\varphi(\bar{r})\big] = -\rho(\bar{r}) , \qquad (3)$$

where $\bar{r} = (x,y,z)$ is a position vector, $\rho$ is the total charge of ions and channel proteins, $\varphi$ is the potential of the electric field, $\varepsilon = \varepsilon_o \varepsilon_r$ - the electric permittivity (where $\varepsilon_o$ is the permittivity of free space, and $\varepsilon_r$ is a relative permittivity) and $\nabla = (\frac{\partial}{\partial x}, \frac{\partial}{\partial y}, \frac{\partial}{\partial z})$ – the gradient operator. One adds to the obtained electric field, an exterior field connected with the membrane potential.

In a very narrow selective channel, in an axon of a nerve cell for example, there can be only few ions of the same type. In such membranes, ions of different types are separated and their transport takes place in different channels. The electric field inside such channels is a sum of fields coming from surface charges and individual ions. The first field is an exterior field in the Langevin equations (2). The second one is a result of interactions between ions. The Langevin equations (2) were derived under the assumption that random Brownian motion of ions is independent of the presence of other ions. The problem of interaction between ions in very narrow channels requires a separate discussion.

In the first paper (Cooper et al. 1985), where Brownian dynamics was used to describe ion transport, one assumed that ions movement is one-dimensional. Such assumption is far from realistic models of ion channels. Papers published since 1998, mainly by a group of physicists from the Australian National University in Canberra, contain numerical simulations in a three-dimensional space. A detailed list of references devoted to this issue can be found in extensive reviews (Chung & Kuyucak 2002; Chung & Tieleman 2006, Kuyucak et al. 2001, Kuyucak & Chung 2002, Kuyucak & Batsug 2003; Roux 2005).

Parameters required in equations (such as an electric permittivity or a diffusion coefficient) are taken from molecular dynamics. One very often uses parameter fitting based on an optimization principle (Corry et al. 2001; Edwards et al. 2002; Mamonov et al. 2003; Mamonov et al. 2006). Geometrical dimensions of channels are often taken from experimentally obtained images or from molecular dynamics. In a recent paper (Krishnamurthy & Chung 2007), a stochastic optimization algorithms were constructed to estimate certain structural parameters of ion channels.

In Brownian dynamics, Langevin equations are solved to describe trajectories of all ions. In order to do so in very short time intervals (steps) of few femtoseconds ($10^{-15}$ s), Langevin equations are integrated to find velocities and locations of all ions before the next step. This procedure is repeated for a sufficiently long time, usually few microseconds ($10^{-6}$ s), to find the number of ions ($\Delta n$) passing through the channel in time ($\Delta t$). This gives us the flux of ions,

$$J = \frac{\Delta n}{\Delta t} \, . \tag{4}$$

Details of this method can be found in a review paper (Kuyucak et al. 2001). Fluxes obtained for different values of a membrane potential for a fixed ion concentration or for different concentrations but a fixed membrane potential, allow us to describe current-voltage and current-concentration characteristics. Such characteristics can be confronted with an experimental data.

In the introduction to (Corry et al. 2001), the authors declare that in their model, based on Brownian dynamics, ion chemical potentials and other channel parameters were neither assumed *ad hoc* or fitted to experimental data. It seems that they cannot really justify it. An optimization of channel parameters and their calculations based on molecular dynamics are not free from necessary approximations (for example treating water in a channel as a continuum, putting the relative electric permittivity of a channel protein to 2 or treating channel proteins as static structures).

Despite many simplifications, the description of ion transport in ion channels based on Brownian dynamics explains more experimentally observed channel characteristics than the PNP theory (Edwards et al. 2002). In particular, it predicts, in the agreement with experiments, current-voltage and current-concentration characteristics (including the observed flux saturation with respect to the ion concentration in a solution near the membrane (LeMasurier et al. 2001; Miller 1999), caused by the independence of the time of the ion passage through a selective filter of the concentration (Berneche & Roux 2003; Chung & Kuyucak 2002; Kuyucak et al. 2001; Kuyucak & Chung 2002; Roux et al. 2004).

A description of the ion transport based on the fenomenological Brownian dynamics is still a rough approximation and does not explain all experimentally observed ion channel characteristics (like channel selectivity for univalent cations).

When we realize that proteins form dynamical structures whose pores, allowing ion transport, have cross-sections of atomic sizes, then we understand that using statistical macroscopic parameters (like an electric permittivity $\varepsilon$, and a diffusion coefficient D inside the channel) to describe their functions is not justified. This may constitute a fundamental limit of usability of Brownian dynamics to describe the ion transport in channels. Moreover, it is argued in (Burykin et al. 2002; Schutz & Warshel 2001) that one cannot define an electric permittivity of protein molecules and a solution near their surface of contact. In particular, one cannot characterize a protein molecule by an average permittivity (an estimated permittivity varies in space and it depends on the method of calculation). Protein channels are commonly treated as equilibrium structures with a time-independent permittivity. In reality, ion channels are non-equilibrium structures in which moving ions induce a time-dependent electric permittivity of channel proteins (Burykin et al. 2002). The above described problem of the electric permittivity of ion channels concerns in the same degree models of Brownian dynamics, molecular dynamics (Murzyn 2002), and the PNP theory.

We would like to pinpoint still another problem. Brownian dynamics describes movements of individual ions. Results of experimental papers (Morais-Cabral et al. 2001) and of molecular dynamics (Berneche & Roux 2000; Berneche & Roux 2003; Shrivastava & Sansom 2000) suggest that potassium ions are moving collectively together with a water molecule between them – they pass from S1 and S2 centers to S2 and S4. Such collective transport cannot be described in a simple way within models based on molecular dynamics. A kinetic theory of collective transport was presented by Nelson (2002, 2003a, 2003b). He assumed that transport barriers exist at the channel entrance and exit and there no barriers inside the channel (which is consistent with molecular dynamics). Results of the above papers display a saturation in current-concentration characteristics. We would also like to point out that ion sizes are compared to those of water molecules, therefore treating ions in channels as Brownian particles is not justified (Graf et al. 2004).

We would like to discuss yet another issue. In many physiology and biophysics textbooks, in chapters devoted to nerve impulses, one uses the concept of permeability of the membrane of a nerve cell for ions of the i-th type. In particular, the permeability constant, $P_i$, is used to estimate the selectivity of ion channels (Hille 2001; Loughed et al. 2004; Tieleman et al. 2003). The time dependence of the action potential is explained by changes of membrane-potential dependent permeability constants of sodium ions, $P_{Na}$, and potassium ions, $P_K$, according to the Goldman-Hodgkin-Katz equation for the diffusive rest potential. One has to remember that the formula derived by Goldman (1943) holds for uniform membranes, where ions are subject to an electro-diffusion through the whole membrane surface. Permeability constants, $P_i$, which appear in his formula are defined by the following expression:

$$P_i = \frac{K_i\, D_i}{d},$$  (5)

where $K_i$ is a coefficient of a ion division between a membrane and an electrolyte, $D_i$ is a diffusion coefficient, and $d$ - the thickness of the membrane.

Because ions in excitable membranes pass through membrane potential-dependent channels, the concept of the constant membrane permeability losses its sense for single channels.

Even if parameters appearing in the right-hand side of (5) are well defined, they should be space-dependent because the channel cross-section varies in space. Moreover, because ion concentrations are not well defined so is the coefficient of the ion division, $K_i$.

Ions conductivities are measurable. They appear in the equation for the current through the membrane in an excited state (Hodgkin et al. 1952). The conductivity of ions of the i-th type is given by the following expression:

$$G_i = n_i \, P_i \, g_i, \tag{6}$$

where $n_i$ is the number of channels of the i-th type on the membrane surface (for example in the case of squid *Loligo* axon, there are 360 sodium channels and 80 potassium ones per one micrometer square), $P_i$ is a membrane potential dependent probability of the channel opening, and $g_i$ is the conductivity of a single channel. Time changes of sodium and potassium ion conductance describe time dependence of the action potential.

Finally, we would like to emphasize once again the fundamental problem present both in continuum models of the PNP type as well as in the Brownian dynamics. In channels containing several ions simultaneously, it seems to be essential to take into account interactions between them. In Langevin equations, random collisions between ions and water molecules and the surface of the channel are represented by a sum of a deterministic friction force and a purely random force. Relations between friction and fluctuations of random interactions are described by a dissipation-fluctuation theory. Such theory requires the system to be in a thermodynamic equilibrium and particles not to interact. Then the diffusion coefficient (which measures the size of fluctuations) is given by the Einstein relation, $D_i = kT/m_i \gamma_i$, where T is the temperature of the system. However, if we take into account ion interactions, then random forces acting on ions are no longer independent. Also the division of a force into a deterministic and a random part becomes problematic.

In (Schuss et al. 2001), authors claim channel solutions are very diluted and therefore one can neglect correlations between random forces exerted on particular ions (however, they explicitly consider interactions between ions). Then they introduce appropriate electro-diffusion equations with a self-consistent electric field – a solution of Poisson equations. Let us observe, however, that solutions in channels, in comparison to diluted surrounding solutions where ions are far apart, are not diluted. Despite that fact that there are only few ions in the channel, due to atomic sizes of channels, ions are close to each other and therefore their interactions cannot be neglected. Moreover, the concept of a self-consistent field treats concentration as a statistical quantity. In the case of a low number of ions, concentration fluctuations are of the order of the concentration itself.

Different approach is contained in (Canales & Sese 1998). The authors analyze there a motion of interacting ions in electrolyte solutions. In the appropriate Langevin equation, a friction force has the form of an integral dependent upon a history. Friction with a memory is also used if sizes of Brownian particles are compared to sizes of solution particles (Kubo 1966) which takes place in ion channels.

Taking into consideration interactions between ions in the description of ion transport in channels requires further studies. Ion interactions cause dependence between fluxes of different ion types. However, interactions between ions of the same type within one channel does not lead to interactions between channels. This was used in (Burykin et al. 2002) where an additivity principle was invoked.

In conclusion, one can say that only a truly microscopic description of ion channels (based for example on molecular dynamics) can fully explain mechanisms of their functioning (Chung & Tieleman 2006). Further development of computational techniques, a more detailed knowledge of a molecular structure of ion channels and also advances of physics of nano-systems are needed to achieve this goal.

**Appendix A**

**Electro-diffusion equations**

Within the Goldman theory (Goldman 1943), description of ion transport through cell membranes is based on the one-dimensional Nernst-Planck (NP) electro-diffusion equations for ion fluxes,

$$J_i = -D_i \left( \frac{dc_i}{dx} + \frac{q_i c_i}{kT} \frac{d\varphi}{dx} \right) \tag{A1}$$

and the Poisson equation,

$$\varepsilon \frac{d^2 \varphi}{dx^2} = \sum_i c(x)_i q_i + \overline{N}(x), \tag{A2}$$

where $D_i$ is a diffusion coefficient for ions of the i-th type, $c_i$ is their concentration, k – the Boltzmann constant, T – the absolute temperature, $q_i$ – a charge of a ion of the i-th type, $\varphi$ – an electric potential, $\varepsilon = \varepsilon_o \varepsilon_r$ - electric permittivity, and $\overline{N}$ - a fixed charge of the membrane. The NP equation can be obtained by statistical considerations (Appendix B).

Taking into account interactions between ions and the membrane requires the presence, in the equation for ion fluxes, $J_i$ , an additional term representing the potential of such interactions. In this case, the generalized NP equation has the following form:

$$J_i = -D_i\left(\frac{dc_i}{dx} + \frac{c_i}{kT}\frac{dU_i}{dx}\right),$$ (A3)

where $U_i = q_i\varphi + W_i$, $q_i\varphi$ is the energy of a ion in the electric field, and $W_i$ – the energy of a non-electrostatic ion-channel interaction (Kurnikova et al. 1999; Levitt 1999; Oosting 1977; Roux et al. 2004). For wide multi-ion channels, of diameters exceeding several times the Debye radius, one can neglect $W_i$. One can also justify for such channels the concept of ions concentration.

If we assume that the electric field is constant in the channel, then the right-hand side of (A2) is zero (electroneutrality). In the stationary state, (A1) can be integrated. One then obtains the following expression for the flux $J_i$:

$$J_i = \frac{\dfrac{D_i q_i}{kT}\dfrac{V_m}{d}\left[c_{i,w} - c_{i,o}\exp\left(-\dfrac{q_i V_m}{kT}\right)\right]}{\exp\left(-\dfrac{q_i V_m}{kT}\right) - 1},$$ (A4)

where $d$ is the thickness of the layer, $V_m$ – a membrane potential, $c_{i,w}$ – ion concentration in the membrane on the boundary with an internal solution, and $c_{i,o}$ – ion concentration in membrane on the boundary with an external solution.

Assuming that an electric current in the membrane is caused by the passive transport of $Na^+$, $K^+$, and $Cl^-$ ions, in the stationary state of the cell (the zero total current), the equation (A4) allows to express the membrane potential $V_m$ in the following Goldman-Hodgkin-Katz form:

$$V_m = \frac{RT}{F}\frac{P_K\left[K^+\right]_w + P_{Na}\left[Na^+\right]_w + P_{Cl}\left[Cl^-\right]_o}{P_K\left[K^+\right]_o + P_{Na}\left[Na^+\right]_o + P_{Cl}\left[Cl^-\right]_w},$$ (A5)

where $P_K$, $P_{Na}$, $P_{Cl}$ are defined above permeability constants for corresponding ions, F is the Faraday constant, R – the gas constant, and square brackets denote mol concentrations of corresponding ions in the interior of the cell (w) and outside it (o).

## Appendix B

## Smoluchowski equation

The Nernst-Planck equation can be obtained from statistical considerations following Einstein and Smoluchowski theory of the Brownian motion. One assumes that ions passing through a channel are subject to a severe dumping which justifies neglecting the inertial term ($m\dfrac{dv}{dt} = 0$) in Langevin equations (2). One obtains reduced Langevin equations,

$$m_j \gamma_j v_j = F_r(t) + q_j E + F_s,$$  (B1)

where the ion velocity is proportional to the total force acting on the ion. Of course, one also has to assume, for (B1) to have a sense, that exterior forces do not vary substantially during dumping. This assumption is dubious in narrow channels – a cross section of such channels is not a constant and therefore ion-channel interactions vary on short distances.

Ions, considered as Brownian particles, move chaotically and their trajectories are stochastic. Their positions and velocities can be described only with certain probabilities which satisfy Fokker-Planck equations. In the dumping case, the probability density of finding a ion of the i-th type at a certain location x at the time t, $p_i(x,t)$, is a solution of the Smoluchowski equation (which is a particular case of the Fokker-Planck equation),

$$\frac{\partial p_i(x,t)}{\partial t} = \frac{\partial}{\partial x}\left[\frac{kT}{\gamma_i}\frac{\partial p_i}{\partial x} - \frac{F_i}{\gamma_i}p_i\right],$$  (B2)

where $F_i = q_i E + F_s$.

Continuity equation for the probability $p_i$ reads

$$\frac{\partial p_i}{\partial t} = -\frac{\partial}{\partial x}J_{p_i}$$  (B3)

From (B2) and (B3) we get the following formulae for probability fluxes:

$$J_{p_i} = -\frac{kT}{\gamma_i}\frac{\partial p_i}{\partial x} + \frac{F_i}{\gamma_i}p_i$$  (B4)

For wide multi-ion channels, their interior can be treated as a continuous media and therefore probabilities $p_i$ can be then replaced by ion concentrations $c_i$. We can also neglect $F_s$ and obtain the following expression for ion fluxes:

$$J_i = -\frac{kT}{\gamma_i}\frac{dc_i}{dx} - \frac{q_i c_i}{\gamma_i}\frac{d\varphi}{dx}$$  (B5)

From the dissipation-fluctuation theory we get the relation between $\gamma_i$ and the diffusion coefficient $D_i$.

$$D_i = \frac{kT}{\gamma_i},$$  (B6)

hence

$$J_i = -D_i\left[\frac{dc_i}{dx} + \frac{q_i c_i}{kT}\frac{d\varphi}{dx}\right]$$  (B7)

For uniform channels, $D_i$ does not depend on $x$ and we get (A1) which is a fundamental equation leading to the Goldman-Hodgkin-Katz equation (Appendix A).